\documentclass[preprint,12pt]{elsarticle}

\usepackage{amssymb}
\usepackage{amsmath}
\usepackage{amsthm}

\usepackage{lineno}

\theoremstyle{plain}
\newtheorem{thm}{Theorem}
\newtheorem{proposition}[thm]{Proposition}

\newcommand{\SC}{\text{S}} 
\newcommand{\LC}{\text{L}} 
\newcommand{\SNb}{\sharp_{\SC}}
\newcommand{\LNb}{\sharp_{\LC}}

\newcommand{\SG}{\sigma} 
\newcommand{\LG}{\lambda} 
\newcommand{\step}{\curvearrowright}

\begin{document}

\begin{frontmatter}


\author{Burghard Herrmann\corref{cor1}}
\ead{burghard.herrmann@fu-berlin.de}


\title{Axioms of Phyllotaxis} 



\affiliation{organization={Freie Universit\"at Berlin, Fachbereiche Mathematik/Informatik und Physik},
            addressline={Arnimallee~14}, 
            city={14195~Berlin},
            state={+49 163 8543044}, 
	country={Germany}
	  }

\begin{abstract}
Observations of simultaneous occurrence of new objects in the early developmental stage of flower heads challenge the classical model of phyllotaxis. Prusinkiewicz (2026) applies Lindenmayer's axiomatic approach to a model of the initial expansion stage and explains the origin of the divergence angle 137.5°, a characteristic by which the classical model consistently demonstrates its validity. This paper proposes a generalised axiomatic model of initial expansion and applies it to a handful of relevant divergence angles. 
\end{abstract}




\end{frontmatter}



\section{Introduction}

This work is inspired by \citet{Turing}, which includes Turing's major unfinished work 'Morphogen Theory of Phyllotaxis', \citet{Gerbera} and \citet{emergence}.

We investigate how a simple form of step-by-step development, described by a non-deterministic Lindenmayer system, realigns an initial structure of phyllotaxis. The initial structure is modeled by a proper binary word, called "the axiom", which represents a flexible partition of an unrolled circle. The main result states that the axiom's partition is realigned by the ratio of subsequent Fibonacci numbers, which converges to the golden ratio. 

\section{Results}

We use the letters $a,b,n,k$ for non-negative integers and $x,y,r$ as well as some greek letters for real numbers. The elements of the Fibonacci sequence $0,1,1,2,3,5,8,\ldots$ are denoted by $F_n$, starting from $F_0=0$ and $F_1=1$. Put $\psi={(\sqrt{5}-1)}/{2}$ for the reciprocal of the golden ratio. 

The unit interval $[0,1]$ is considered as unrolled circle with unit circumference. A value $x\in[0,1]$  represents an angle of $x\times 360$\textdegree. Conversely, an angle divided by {$360$\textdegree} represents the length of a subinterval of $[0,1]$, e.g., $137.5/360$ approximates $1-\psi$.

Throughout this paper we work with binary words by means of strings of characters {\SC} and {\LC}. We use the letters $v,w$ for words and $C$ for characters.

Let $w=C_1\ldots C_k$ be a  binary word where $k$ is the length of $w$.
The count of $\SC$ and $\LC$ in $w$ is denoted by $\SNb(w)$ and $\LNb(w)$, respectively.  Of course, $\SNb(w)+\LNb(w)=k$.
If $\SNb(w)$ and $\LNb(w)$ both are positive, $w$ is called \textit{proper binary}. 

\subsection{Flexible partitions}
\label{subsecP}
Let $(a,b)$ be a pair of positive integers. We consider pairs of real numbers $0\leq x\leq y$ such that 
\begin{equation}
\label{eqLin}
ax+by=1.
\end{equation}
They are uniquely determined by their ratio $r=x/y\in[0,1]$ via $x=ry$ and $y=1/(ra+b)$.

Let $w=C_1\ldots C_k$ be a proper binary word with $\SNb(w)=a$ and $\LNb(w)=b$.
For  a ratio $r\in[0,1]$ put
\begin{align*}
\LG(w,r) & :=1/(ra+b)  \text{ and } \\
\SG(w,r) &:=r\LG(w,r).
\end{align*}
Note that, $0\leq\SG(w,r)\leq\LG(w,r)$ is the unique solution of (\ref{eqLin}) with ratio $r$.
Hence, there is a partition of the unit interval into $k$ subintervals such that the length of the $j$-th subinterval is $\SG(w,r)$ if $C_j=\SC$ and $\LG(w,r)$ if $C_j=\LC$ for $j=1,\ldots,k$. This partition is denoted by $P(w,r)$. Taking $r\in[0,1]$ as variable means that we interprete $w$ as a flexible partition.

\subsection{Simple expansion}

Let $w=C_1\ldots C_k$ be a  binary word. We say that $w'$ \emph{derives from $w$ by simple expansion}, in symbols $w\step w'$, if  
\begin{equation}\label{ambE}
C_j\text{ is replaced by } 
\begin{cases} \LC & \text{ if } C_j=\SC,\\
{\SC\LC} \text{ or } {\LC\SC} & \text{ if } C_j=\LC,
\end{cases} 
\end{equation}
for $j=1,\ldots, k$, i.e., the letters of $w$ are simultaneously replaced.
We write $w\step^n w'$ if $w'$ derives from $w$ by $n$ steps of simple expansion. More precisely, $w\step^n w'$ means that there are $w_0,\ldots,w_{n}$ such that
$$
w=w_0 \step w_1 \step \ldots \step w_n=w'. 
$$
For example, {\LC\SC\LC} derives from {\LC} in two steps of simple expansion. Note that there are two different derivations:
$ \LC \step{}\LC\SC\step{}\LC\SC\LC, $ and 
$ \LC \step{}\SC\LC\step{}\LC\SC\LC. $
However, we have
\begin{equation}\label{seedNb}
C\step^n v \text{ implies }(\SNb(v),\LNb(v))= 
\begin{cases} 
(F_{n-1},F_n) & \text{ if } C=\SC,\\
(F_n,F_{n+1})& \text{ if } C=\LC.
\end{cases} 
\end{equation}
By (\ref{ambE}),  the latter is clear for $n=1$ and proceeds by induction for $n\geq2$.

In the terminology of Lindenmayer systems, $(\{\SC,\LC\},w,R)$ is a non-deter-ministic L-system with axiom $w$ where $R$ is the set of rules that corresponds to (\ref{ambE}). 

\subsection{Aggregation of refined partitions}
Let $w$ be a proper binary word and $r\in[0,1]$. Our main result concerns the partition $P(w,f_n(r))$ where $f_n(r)$ is defined as follows 
\begin{equation*}
f_n(r):=\frac{F_{n-1}r+F_{n}}{F_{n}r+F_{n+1}}.
\end{equation*}
As is easiliy seen, $f_n(r)$ is strictly monotone for $r\in[0,1]$. As $f_n(0)={F_n}/{F_{n+1}}$ and $f_n(1)={F_{n+1}}/{F_{n+2}}$, 
\begin{equation}\label{nested}
f_n([0,1]) \text{ ranges between } {F_{n}}/{F_{n+1}} \text{  and } {F_{n+1}}/{F_{n+2}}.
\end{equation}

\begin{proposition}\label{proRatio}
Let $w$ be a proper binary word,  $w\step^n w'$, and $r\in[0,1]$.  Then aggregation of $P(w',r)$ provides $P(w,r')$ where $r'=f_n(r)$.  
\end{proposition}
\begin{proof} 
Let $w=C_1\ldots C_k$ and $j_0, j_1$ such that $C_{j_0}=\SC$ and $C_{j_1}=\LC$. We assume $w\step^n w'$. Thus, there are subwords $v_j$, $j=1,\ldots,k$,  
satisfying $C_j\step^n v_j$ and $w'=v_1\ldots v_k$.  

For $r\in[0,1]$ put $\mu(\SC)=\SG(w',r)$ and $\mu(\LC)=\LG(w',r)$ which are the lengths of the subintervals of $P(w',r)$. Aggregation of $P(w',r)$ means 
\begin{equation*}
\mu(v)=\SNb(v)\mu(\SC)+\LNb(v)\mu(\LC)
\end{equation*}
for any binary word $v$. In particular, $\mu(w')=1$.
By (\ref{seedNb}) and $\mu(\SC)=r\mu(\LC)$, it follows
\begin{equation*}
\mu(v_j)= 
\begin{cases} 
(F_{n-1}r+F_n) \mu(\LC) & \text{ if } C_j=\SC,\\
(F_n r+F_{n+1}) \mu(\LC) & \text{ if } C_j=\LC.
\end{cases} 
\end{equation*}
For $a=\SNb(w)$ and $b=\LNb(w)$, we conclude that $0\leq\mu(v_{j_0})\leq\mu(v_{j_1})$ solves (\ref{eqLin}) with ratio $f_n(r)$. This completes the proof.
\end{proof} 


\section{Applications}
As is well-known, $\lim_{n\rightarrow\infty}F_{n}/F_{n+1}=\psi$. By (\ref{nested}), $\lim_{n\rightarrow\infty}f_n([0,1])=\{\psi\}$.
Thus, for a proper binary word $w$, the partitions $P(w,f_n(r))$ provided by Proposition~\ref{proRatio} converge to $P(w,\psi)$. 

Finally, we provide a complete list of axioms $w$ with $a=\SNb(w)$, $b=\LNb(w)$, $a+b\leq5$, small angles $\psi/(\psi a+b)\times$360{\textdegree} and divergence angle less than 180\textdegree:
 
\begin{tabular}{lrrr}
\textbf{axiom} $w$ &$(a, b)$  & \textbf{small angle} &  \textbf{divergence}
\\
\SC\LC&(1,1) &137.5\textdegree  &  137.5\textdegree 
\\
\SC\SC\LC&(2,1) &99.5\textdegree &  99.5\textdegree
\\
\SC\SC\SC\LC&(3,1)&78.0\textdegree&   78.0\textdegree
\\
 \SC\SC\SC\SC\LC&(4,1)&64.1\textdegree&  64.1\textdegree
\\
 \LC\SC\LC\SC\SC&(3,2)&57.7\textdegree&  151.1\textdegree
\end{tabular}

\noindent
where for $w=\LC\SC\LC\SC\SC$ the large angle of $P(w,\psi)$  is 93.4{\textdegree} and the divergence is the sum of a small and a  large angle. 

\appendix




\section*{Acknowledgement}

Many thanks to my family as well as participants of workshops at the Max Planck Institute for Plant Breeding Research, the 22nd Int. Conf. on Fibonacci Numbers and Their Applications,  and Colloquium Logicum 2026.

\end{document}